\documentclass[aps,prl,twocolumn,showpacs,letterpaper,preprintnumbers,amsmath,amssymb]{revtex4}

\usepackage{graphicx}
\begin{document}


\title{Nondestructive Fluorescent State Detection of Single Neutral Atom Qubits}


\author{Michael J. Gibbons}
\author{Christopher D. Hamley}
\author{Chung-Yu Shih}
\author{Michael S. Chapman}
\affiliation{School of Physics, Georgia Institute of Technology,
  Atlanta, GA 30332-0430 }

\date{\today}

\begin{abstract}

 We demonstrate non-destructive (loss-less) fluorescent state detection of individual neutral atom qubits trapped in an optical lattice.  The hyperfine state of the atom is measured with a 95\% accuracy and an atom loss rate of 1\%.  Individual atoms are initialized and detected over 100 times before being lost from the trap, representing a 100-fold improvement in data collection rates over previous experiments.  Microwave Rabi oscillations are observed with repeated measurements of one-and-the-same single atom.

\end{abstract}

\pacs{03.67.-a, 37.10.Gh, 42.50.Ct}

\maketitle

The development of techniques to trap individual laser cooled atoms and ions has led to frequency metrology of unprecedented accuracy \cite{wine} and have enabled pioneering experiments in the field of quantum information processing \cite{blatt, monroe, saffman}. The success of these systems is due to the isolation of the atom from external environmental perturbations and the facility with which the quantum states of the atom can be initialized, manipulated and detected using lasers and other electromagnetic fields.

Quantum state readout in ion traps has largely been done by direct detection of state-selective fluorescence, first used to observe quantum jumps in atomic systems \cite{dehmelt, berquist, sauter}. Efficient state detection requires scattering 100s of photons from the atom for typical fluorescence collection efficiencies of $\sim$1\%. Each scattering event heats the atom by an amount comparable to the recoil temperature $T_{recoil} = \hbar ^{2} k^{2} / mk_{B}$. For trapped ions, this heating is negligible compared with the large depth ($>$1000 K) of the traps and hence quantum state readout using direct detection of state-selective fluorescence can be achieved with no loss of the ions.

Neutral atom traps are much shallower, typically $\sim$1 mK, and hence the heating induced by fluorescence state detection can be comparable to or exceed the depth of the trap. As an alternative, state-selective ejection of atoms was developed for accurate quantum state measurement of individually trapped neutral atom qubits \cite{kuhr1, kuhr2, yavuz, volz, lengwenus, jones}. In this technique, rather than trying to minimize the atom heating, the atoms in one quantum state are deliberately heated out of the trap with strong, unbalanced radiation pressure. Subsequently, the remaining atoms in the quantum register (which are now known to be in the other quantum state) are detected using radiation that is not state selective and is detuned to provide simultaneous cooling of the atoms \cite{kuhr3}.

 While state-selective ejection of neutral atoms is a very effective detection method, the atom traps must be reloaded after every readout operation, which limits the experiments to a $\sim$1 s${}^{-1}$ repetition rate. These limitations will need to be overcome to significantly advance the field of neutral atom quantum information processing. One solution that has been demonstrated is to use a cavity QED system to increase the collection efficiency of the scattered photons \cite{boozer, khudaverdyan, bochmann, gehr}. The quantum state can then be determined with fewer scattering events, resulting in lower heating and minimal loss of the qubits. A drawback of this approach is that cavity QED systems significantly complicate the experimental setup and each atom to be detected needs be localized within the small cavity mode.

 Here, we revisit the prospect of simple fluorescence detection of trapped neutral atoms. In contrast to recent work that concluded that atom loss due to heating would limit fluorescence detection fidelity to $<$50\% in optical traps  \cite{bochmann}, we demonstrate that it is possible to achieve accurate fluorescence state measurement with minimal loss without using cavity enhanced detection. Using a high numerical aperture lens, single ${}^{87}$Rb atom hyperfine qubits are detected with 95\% accuracy and an atom loss rate of $\sim$1\%. With this technique, we measure single atom Rabi flopping using microwave transitions for $\sim$50 state preparation and detection cycles with one-and-the-same atom.

 We begin our discussion with simple estimates to show the feasibility of our approach.  Using ${}^{87}$Rb as an example, we consider qubit states stored in the $F=1$ and $F=2$ hyperfine states separated in energy by 6.8 GHz. Standard detection of this qubit employs excitation of the quasi-cycling, $5S_{1/2} ,F=2\to $$5P_{3/2} ,F'=3$ transition.  In order to achieve lossless quantum state detection with high accuracy, several objectives need to be met. It is necessary to detect enough scattered photons to determine the quantum state of the atom with high fidelity, discriminating against stray photons and background noise of the detector. The transition is only quasi-closed: for resonant excitation of the $F=2\to $$F'=3$ transition, there is a small probability for off-resonant excitation of the $F=2\to $$F'=2$ transition, which can `depump' to the $F=1$ state leading to a detection error.  Finally, the heating of the atom due to the excitation needs to be much less than the depth of the optical trap.

 We first consider excitation for a fixed duration of time such that the mean number of detected photons is $\bar{m}$ for the quantum state $F=2$ and zero for $F=1$, and we ignore depumping and detector noise. In this case, the detection error will be the probability of detecting zero photons when the quantum state is $F=2$, which is $P_{0} =\exp (-\bar{m})$ according to Poissonian statistics.  Hence to achieve an error rate $<1\% $ requires $\bar{m}\ge 5$ and an error rate $<0.1\% $ requires $\bar{m}\ge 7$.

 It is possible to achieve a net photon collection + detection efficiency of 2\% using an off-the shelf large-numerical-aperture objective (NA = 0.4) and a single photon avalanche photodiode counting module (SPCM) with $\sim$50\% quantum efficiency at $\lambda $ = 780 nm.  Thus, for a state detection error rate of $<$1\%, the atom must scatter 250 photons. The heating of the atom can be estimated by considering that each absorption-emission cycle heats the atom approximately 2\textit{T${}_{recoil}$} = 720 nK, which yields a total heating of 180 $\mu$K for 250 scattered photons. This heating is much less than the 1-2 mK trap depth typically used for neutral atom qubit optical traps, and hence the chance of ejecting the atom is small.

 The error level due to accidental depumping depends critically on the detuning of the probe beam. Off-resonant excitation of the $F'=2$ level scales as $(\gamma /2\Delta _{2'-3'} )^{2} $ relative to resonant excitation of the $F'=3$ level, where $\gamma $ = 6 MHz is the linewidth of the excited state and $\Delta _{2'-3'} $= 266 MHz is the energy difference between the $F'=2$ and $F'=3$ states.  If the probe beam is exactly on resonance, then off-resonant excitation is suppressed by a factor of $\sim$8000. However, at a probe detuning equal to the linewidth of the transition, this value drops to 1600, and at two linewidths it drops to only 450. Tuning the probe laser exactly on resonance is not difficult; however, the differential AC Stark shift of the transition due to the optical trapping fields can result in effective detunings comparable to the transition linewidth and hence must be considered.

 It is possible to do better than the estimates above by probing only until a predetermined number of photons, \textit{N${}_{D}$}, is detected instead of probing for a fixed duration. This removes the statistical uncertainty related to the detection of a mean number of photons. In the absence of background scatter or detector noise, it is necessary to detect only a single photon from the $F'=3$ state in order to determine that the qubit was in the $F=2$ state with no error. However, the photon counter has a dark count rate of $\sim$100 counts/s so the probability of receiving a false detection even during a short ($<$1 ms) pulse is appreciable. In practice therefore, choosing \textit{N${}_{D}$} = 2 insures that the signal is due to an atom in the $F=2$ state with detector noise error rate of 10${}^{-4}$ for a 1 ms maximum pulse length.

 We now turn to the experiment, where we describe the basic experimental set-up (for additional details, see \cite{gibbons}) and procedure, and present results demonstrating the proof of principle.  The experiment begins with a magneto-optic trap (MOT) operated in the single atom regime \cite{hu}. In order to load single or small numbers of atoms, the MOT is operated at magnetic field gradients of $\sim$250 G/cm to decrease the loading volume.  This also provides tight confinement of the atoms, localizing them to a diameter of approximately 25 $\mu$m. The atoms are detected and counted by measuring the fluorescence of the atoms from the MOT cooling beams. The atoms are captured in an evacuated quartz cell from a background pressure $<$10${}^{-}$${}^{11}$ torr using trapping beams with a diameter of 1 mm, an intensity of 10 mW/cm${}^{2}$, and a detuning of $-$10 MHz from resonance. Loading time to trap a single atom is approximately 2 s.

 The trapped atom(s) are transferred to a dipole trap that uses a 1064 nm ytterbium fiber laser beam focused on the MOT, with a waist of 13 $\mu$µm. The beam is retro-reflected to produce an optical lattice with a trap depth of 2 mK.  The optical trap is on during the MOT loading time, and the atoms are transferred to the optical trap by turning off the magnetic field gradient and increasing the cooling beam detunings to $-$20 MHz to optimize continuous cooling and observation of the optically trapped atoms.

 Fluorescence imaging is used to ensure that precisely one atom is confined in the trap. Fluorescence from the atoms is captured by a long working distance microscope objective (NA = 0.4) mounted outside the glass cell. The light passes through a beam splitter that sends 5\% of the light to a CCD camera. The remaining light is focused onto a SPCM used for state detection. The CCD camera takes an image with a 1 s exposure time to determine if a single atom has been loaded within the area of interest (defined by the field of view of the single photon counter) and none elsewhere in the trap. Successful loading occurs slightly less than 1/3 of the time. Fig. 1 shows a single atom loaded within the area of interest.

\begin{figure}
\includegraphics*[width=2.45in, height=0.75in, keepaspectratio=false]{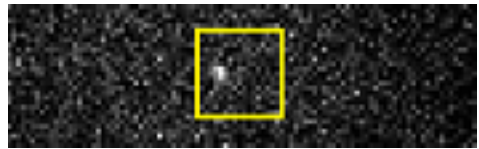}
\caption{Fluorescent image of a single atom acquired with a CCD camera. The field of view of the single photon counter is shown as a yellow square.}
\end{figure}

 Once a single atom is successfully loaded into the region of interest, optical pumping is used to prepare it in the desired quantum state.  For initial experiments, we ignore the Zeeman structure and consider a qubit with states stored in the $F=1$ and $F=2$ hyperfine levels separated in energy by 6.8 GHz. The hyperfine states are prepared in the standard way using a 10 ms pulse of either the MOT cooling lasers or the repump laser which is tuned to the $F=1\to $ $F'=2$ transition.

 Quantum state readout is performed using two 6 $\mu$W counter-propagating probe beams, focused to 125 $\mu$m and detuned +5 MHz from $F=2\to $$F'=3$ transition. For a fixed probe time of 300 $\mu$s, the average number of collected photons from the $F=2$ state is measured to be $\bar{m}=21$ and from the $F=1$ state is $\bar{m}=0.3.$ Following the discussion above, for best performance, the output of the photon counter is monitored in real time as the atom is being probed for up to 300 $\mu$s, and as soon as two counts have been received, the atom is determined to be in the $F=2$ state and the probe beam is extinguished.

 Once the state of the atom has been determined, the cooling lasers are turned back on, and the CCD camera takes another picture of the trap to determine whether the atom has remained trapped. The distribution of total counts on the single photon counter for atoms prepared in the $F=1$ and $F=2$ states is shown in Fig. 2 for over 1600 trials.

\begin{figure}
\includegraphics*[width=3.26in, height=1.72in, keepaspectratio=false]{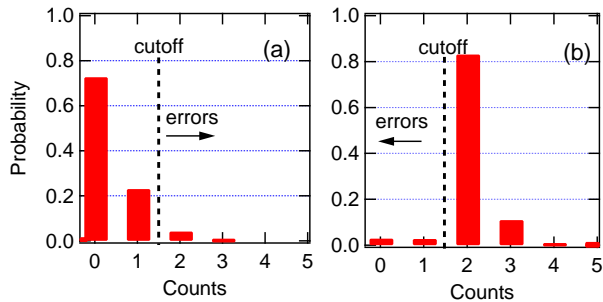}
\caption{Histogram of counts per atom, where the probe was extinguished after the photon detector recorded two counts. (a) Atoms were prepared in the $F=1$ hyperfine state.  Any signal above one count represents an error. (b) Atoms were prepared in the $F=2$ hyperfine ground state. Any signal below two counts represents an error.}

\end{figure}

Two types of errors are possible. We define the `$F=1$ error rate' as the probability of falsely detecting the atom in the \textit{F} = 2 state when it was prepared in the $F=1$ state. Atoms in the $F=1$ state are not affected by the probe beam, so no counts are expected apart from background noise. We define the `\textit{F} = 2 error rate' as the probability of failing to detect an atom that was prepared in the \textit{F} = 2 state. The data in the histogram gives error rates of 4\% and 5.5\% for $F=1$ and \textit{F} = 2 respectively. The measured atom loss rates are 0.9\% (1.05\%) for state preparation and detection in the $F=1$$(F=2)$state, respectively. These results are shown graphically in Fig. 3.

\begin{figure}
\includegraphics*[width=2.91in, height=1.57in, keepaspectratio=false]{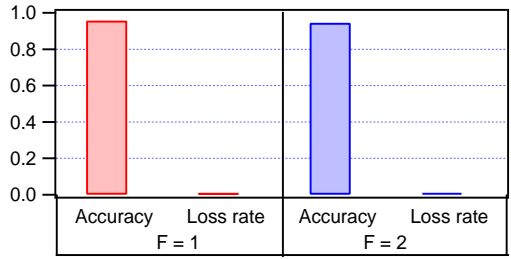}
\caption{Accuracy and loss rate for detection of the different hyperfine states. For the \textit{F} = 1 initial state (1684 data points), the accuracy was 96\% and the loss rate was 0.9\%. For the \textit{F} = 2 initial state (2127 data points), the accuracy was 94.5\% and the loss rate was 1.05\%.}

\end{figure}

The measured loss rates are low enough to enable preparation and detection of each atom many times before losing it from the trap. Fig. 4 shows the results obtained for 100 repeated measurements per atom using a total of 102 individual atoms.  In each of the 102 atoms, the atom undergoes 100 cycles of state preparation to the \textit{F} = 2 state, followed by state-detection.  In order to counter the heating associated with the detection process, a 5 ms cooling pulse is applied to the atom following each state preparation and detection cycle.

\begin{figure}
\includegraphics*[width=3.13in, height=2.89in, keepaspectratio=false]{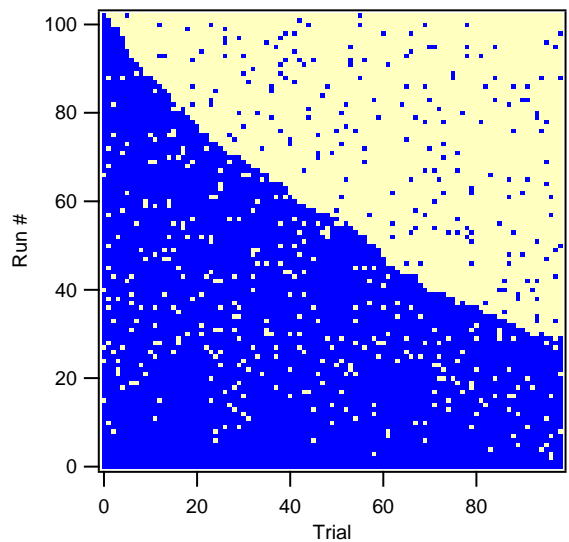}
\caption{102 individual atom runs. For each run, a single atom is prepared in the \textit{F} = 2 state and then detected, repeated for 100 trials. The dark pixels correspond to a positive detection of the atom in the \textit{F} = 2 state. The runs are numbered in order by how long the atom remained in the trap.}
\end{figure}

In Fig. 4, each row corresponds to the series of state measurements of a single atom prepared in the \textit{F} = 2 state and the dots correspond to a positive detection of the atom in the \textit{F} = 2 state. The individual atom runs are sorted in order by how long the atom remained in the trap in order to illustrate the atom loss probability. The missing dots on the lower part of the graph indicate \textit{F} = 2 errors, while the stray dots on the upper part of the graph indicate $F=1$ errors. The shape of the data envelope reveals the atom loss probability. An exponential fit to the average of all of the runs yields an 86 cycle lifetime, which implies a loss rate per cycle of 1.2\%. This matches the previously measured loss rate reasonably well. It is noteworthy that in $\sim$30\% of the cases, the atom survives for the full 100 trials.

State preparation to the \textit{F} = 1 or \textit{F} = 2 quantum states as used above provides a technically expedient method to assess the performance of the quantum state detection method. Of course, for useful applications to quantum information, we are interested in measuring qubits of arbitrary superpositions of the two states. As a first demonstration, we have applied our technique to measure qubits created using microwave rotations between the two hyperfine states, and we have measured Rabi oscillations with one and the same atom.

\begin{figure}
\includegraphics*[width=3.29in, height=3.30in, keepaspectratio=false]{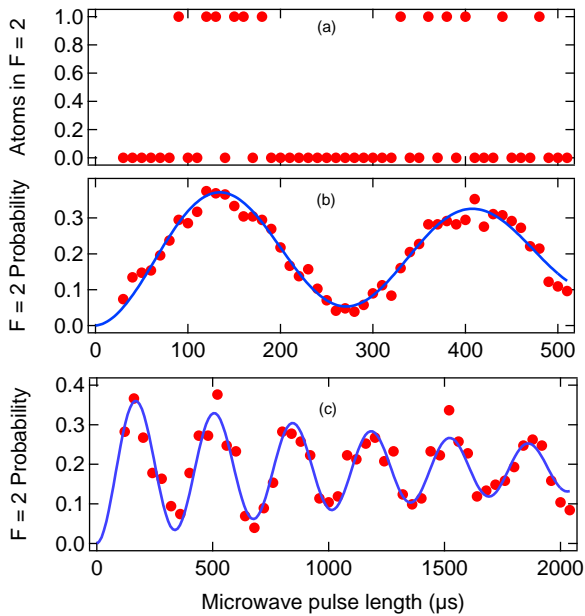}
\caption{Microwave Rabi flopping on the \textit{F} = 1, \textit{m${}_{F}$} = 0 ? \textit{F} = 2, \textit{m${}_{F}$} = 0 `clock' transition, using non-destructive state measurement. (a) Typical data showing a single atom prepared and measured for fifty different microwave pulse lengths. (b) Average data for 312 single atom curves. (c) Similar data for longer microwave pulses.  The solid line indicates a fit to the data, with a Rabi rate of 2.95 kHz and a decoherence time of 2.2 ms.}
\end{figure}

The experimental sequence is very similar to the previous section except that the atom is initialized to the \textit{F }= 1 state and then excited to a superposition of the \textit{F} = 1 and \textit{F} = 2 states using a pulse of microwave radiation tuned to the\textit{ F} = 1 $\rightarrow$ \textit{F} = 2 hyperfine transition. For each atom loaded, fifty cycles are run, with a variable microwave pulse length proportional to the cycle number.

Fig. 5(a) shows the results for a typical single atom and demonstrates Rabi flopping with repeated measurements of a single atom. While the outcome of each individual measurement is either the \textit{F} = 1 or \textit{F} = 2 state, the probability of finding the atom in the \textit{F} = 2 state is periodic in pulse length.  In Fig. 5(b), the data for the average of 312 atoms is shown, together with a sinusoidal fit.  Two Rabi oscillations are clearly observed.  In Fig. 5(c), similar data is shown for longer microwave pulses.  A fit to the data matches a damped Rabi oscillation with a Rabi rate of 2.95 kHz and a decoherence time of 2.2 ms, limited by the differential Stark shift of the dipole trap.

In these data, the maximum probability of finding the atom in the \textit{F} = 2 state is approximately 1/3, due to the multiplicity of the Zeeman states that we have so far ignored. The state preparation to the \textit{F} = 1 initial state should equally populate the three \textit{F} = 1, \textit{m${}_{F}$} = 0, $\pm$1 Zeeman states.  On the other hand, the microwave radiation is tuned to the \textit{F} = 1, \textit{m${}_{F}$} = 0 $\rightarrow$ \textit{F} = 2, \textit{m${}_{F}$} = 0 `clock' transition, which is insensitive to magnetic fields to first order.  The microwave radiation is not resonant with transitions from the \textit{F} = 1, \textit{m${}_{F}$} = $\pm$1 Zeeman states to the \textit{F} = 2 states, due to Zeeman shifts, so these states are not excited.  As a result, we expect the maximum excitation to the \textit{F} = 2 state to be approximately 1/3.

In summary, we have demonstrated nearly lossless quantum state detection of single ${}^{87}$Rb atoms with 95\% accuracy with an atom loss rate of 1\% using fluorescent detection. Individual atoms have been state prepared and detected up to 100 times, and we have measured single atom Rabi flopping using microwave transitions with one-and-the-same atom. While these proof-of-principle demonstrations are already potentially impactful for neutral atom quantum information experiments, we are confident that straightforward extensions of this work will lead to an order of magnitude improvement in our results.  We acknowledge support from the NSF (PHY-0703030).

\end{document}